\documentclass[twocolumn,showpacs,amsmath,amssymb,prb,superscriptaddress]{revtex4}
\usepackage{bm}
\usepackage[normalem]{ulem}
\usepackage{graphicx}
\usepackage{color} % for colored comments

\begin{document}

\title{Thermodynamic nature of the $\mathbf{0-\pi}$ quantum transition in superconductor-ferromagnet-superconductor trilayers}

\author{N. Pompeo}
\author{K. Torokhtii}
\affiliation{Dipartimento di Ingegneria, Universit\`a Roma Tre, Via della Vasca Navale 84, 00146 Roma, Italy}
\author{C. Cirillo}
\affiliation{CNR-SPIN Salerno and Dipartimento di Fisica ``E.~R. Caianiello'', Universit\`{a} di Salerno, I-84084 Fisciano (SA), Italy}
\author{A.~V.~Samokhvalov}
\affiliation{Institute for Physics of Microstructures, Russian
Academy of Sciences, \\ 603950 Nizhny Novgorod, GSP-105, Russia}
\affiliation{Lobachevsky State University of Nizhny Novgorod,
Nizhny Novgorod 603950, Russia}
\author{E.~A. Ilyina}
\author{C. Attanasio}
\affiliation{CNR-SPIN Salerno and Dipartimento di Fisica ``E.~R. Caianiello'', Universit\`{a} di Salerno, I-84084 Fisciano (SA), Italy}
\author{A.~I.~Buzdin} \affiliation{Institut Universitaire
de France and University Bordeaux, LOMA UMR-CNRS 5798, F-33405
Talence Cedex, France}
\author{E. Silva}
\affiliation{Dipartimento di Ingegneria, Universit\`a Roma Tre, Via della Vasca Navale 84, 00146 Roma, Italy}

% 74.25.nn  Surface impedance
% 74.45.+c  Proximity effects; Andreev reflection; SN and SNS junctions
% 74.78.Fk  Multilayers, superlattices, heterostructures

\date{\today}
%
%
%----------- mettere F-layer S-Layer ovunque e controllare 0 e pi coi trattini
%
%

%---------------------------ABSTRACT
\begin{abstract}
In structures made up of alternating superconducting and ferromagnet layers (S/F/S heterostructures), it is known that the macroscopic quantum wavefunction of the ground state changes its phase difference across the F--layer from $0$ to $\pi$ under certain temperature and geometrical conditions, whence the name ``$0-\pi$" for this crossover. We present here a joint experimental and theoretical demonstration that the $0-\pi$ is a true thermodynamic phase transition: microwave measurements of the temperature dependence of the London penetration depth in $\mathrm{Nb/Pd_{0.84}Ni_{0.16}/Nb}$ trilayers reveal a sudden, unusual decrease of the density of the superconducting condensate (square modulus of the macroscopic quantum wavefunction) with decreasing temperature, which is predicted by the theory here developed as a transition from the $0-$ to the $\mathbf{\pi}-$state. Our result for the jump of the amplitude of the order parameter is the first thermodynamic manifestation of such temperature-driven quantum transition.
\end{abstract}

\maketitle
%===============================================================

\section{Introduction}
%---------------------------INTRODUCTION
Superconductor (S)/Ferromagnet (F) structures are the ideal playground for the search for novel phase transitions: two different, competing orderings come into play in a controllable fashion. In fact, S/F heterostructures can be grown with different S and F thicknesses, $d_s$ and $d_f$, respectively, so that the superconducting and ferromagnetic ordering have different effect. Moreover, other external parameters, such as the temperature $T$, can be used to vary the interactions between F and S. In the recent past, the attention has been focussed on the damped oscillatory behaviour of the Cooper pairs superconducting wavefunction $\Delta=|\Delta|e^{i\phi}$ in the ferromagnet:\cite{Buzdin-Bulaevskii-JETPL82,Buzdin-Kuprianov-JETPL91} the peculiarity of the S/F interaction makes the superconducting wavefunction $\Delta$ oscillate in the F--layer, in addition to the conventional, ``proximity-effect-like'', exponential decrease. This unique feature of S/F structures is at the origin of the superconductor-ferromagnet-superconductor (SFS) $\pi$-Josephson junctions (for a review, see Refs. \onlinecite{Buzdin-RMP05,Golubov-RMP04}) characterized by a ground state at phase difference $\delta \phi = \pi$ between S--layers (one can represent this effect by a sign change of the macroscopic superconducting wavefunction across the F--layer). In this case, one speaks of ``$0-\pi$ transition''. A sketch of the spatial dependence of the wavefunction in the $0-$ and $\pi-$ states is reported in Fig.~1: in the ``$0$''  state, $\Delta$ is only depressed in the F--layer. In the ``$\pi$'' state, $\Delta$ has zero value in the middle of the F--layer, and changes sign.
The $\pi$ shift has, among the others, the spectacular consequence of spontaneous supercurrents in ring-shaped structures incorporating $\pi$ junctions.\cite{frolov}
%
%Moreover, the same $\pi$ junctions are at the hearth of a significant shrinking of the size of  high speed superconducting digital single flux quantum circuits, to the extent that the size of superconducting circuits can rival with  traditional semiconductor CMOS-based components\cite{sidorenko}.

Up to now, experiments were directed towards transport observations of the $\pi$--shift of the wavefunction: the efforts were concentrated in measurements of the critical current of Josephson junctions with F barrier, SFS Josephson junctions. In order to avoid excessive depression of the superconducting wavefunction and in order to have well-established superconducting electrodes, the typical structure used to detect the $\pi$--shift was made up of relatively thick S--layers (on the scale of the superconducting coherence length $\xi_s$), and F--layers of thickness of the order of the ferromagnetic coherence length, $\xi_f= \sqrt{\hbar D_f/E_{ex}}=\sqrt{D_f/h}$ (here, $D_f$ is the diffusion coefficient in the ferromagnet, $E_{ex}$ is the exchange energy and $h$ represents the exchange field -- in appropriate units -- acting on the electron's spins\cite{Ryazanov-PRL01,Oboznov-Ryazanov-Buzdin-PRL06}). Since $\xi_f$ is very small (few nanometers) in strong ferromagnets, weak ferromagnets have been often employed for the ease of controlling the geometrical conditions for the $0-\pi$ transition.
The sign change of $\Delta$ has been experimentally observed in SFS Josephson junctions, so that evidence for the $\pi$--shift is nowadays robust: measurements of the critical current of SFS junctions with CuNi alloys as F--barrier\cite{Ryazanov-PRL01,Oboznov-Ryazanov-Buzdin-PRL06} showed $0-\pi$ crossover as a function of $d_f$. Moreover, the subtle role of the temperature $T$ has been revealed: only in the vicinity of the so-called critical thickness condition ($d_f \simeq \xi_f$) can the temperature trigger the $0-\pi$ crossover.\cite{Ryazanov-PRL01,Oboznov-Ryazanov-Buzdin-PRL06} Also, the $\pi$--shift was observed in SFS junctions with PdNi F--barrier\cite{Kontos-PRL02} with $d_f \sim \mathrm{6\,nm}$, even if the transition could not be driven by the temperature in the latter case. So, SFS Josephson junctions with CuNi alloys as F--layer remain the only example of temperature mediated $0-\pi$ crossover.

Aim of this work is to assess the {\em thermodynamic} nature of the $0-\pi$ transition: while the crossover clearly exists in the {\em relative phase between two superconducting electrodes}, no evidence for the change of the {\em order parameter} $|\Delta|^2$ of the macroscopic wavefunction has been brought forward: the finding of a change in $|\Delta|^2$ at the $0-\pi$ transition would be a direct evidence for the thermodynamic nature of such a transition. In particular, the temperature dependence of the amplitude of the order parameter at the critical thickness has not been studied neither theoretically nor experimentally: measurements of the London penetration depth in Nb/Ni bilayers (hybrids based on a strong ferromagnet) as a function of the Ni layer thickness\cite{Lemberger-JAP08} showed only the $d_f$ dependence of the extrapolated zero-temperature superfluid density, similarly to single-$T/T_c$ estimates in Nb/PdNi/Nb structures,\cite{pompeoJS14} with no reported temperature dependence.

We are interested in measuring the temperature dependence of the amplitude of the {\em overall} macroscopic wavefunction. This aim imposes constraints on the geometrical structure under investigation: while in previous experiments\cite{Ryazanov-PRL01,Oboznov-Ryazanov-Buzdin-PRL06,Kontos-PRL02,Sellier-PRB03} the superconducting electrodes were thick enough so that the influence of the $0-\pi$ transition on the S-layers superconducting wavefunction $\Delta$ was negligible, here thin S--layer will be more suited.

%%%%%%%%%%%%%%%%%%%%%%%%%%%%%%%%%%%%%%%%%%%%
\begin{figure}
\includegraphics[width=8 cm]{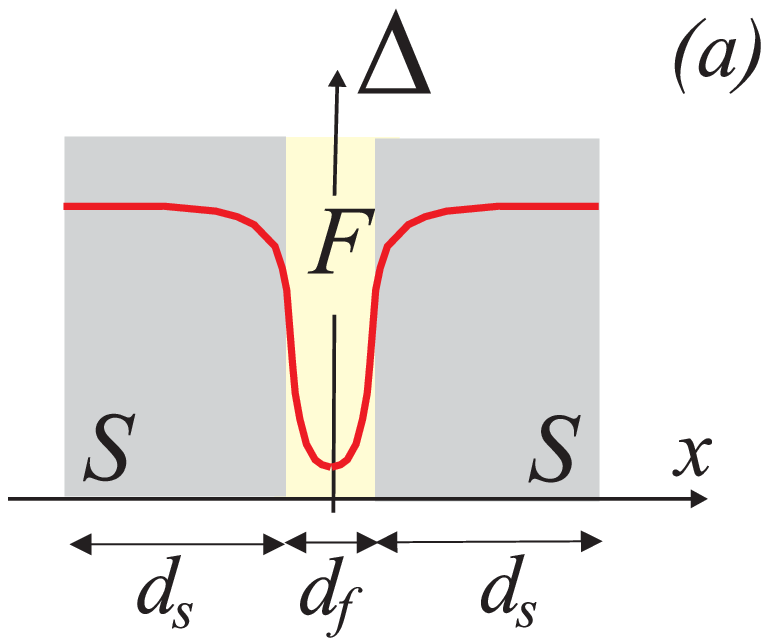}\\
\includegraphics[width=8 cm]{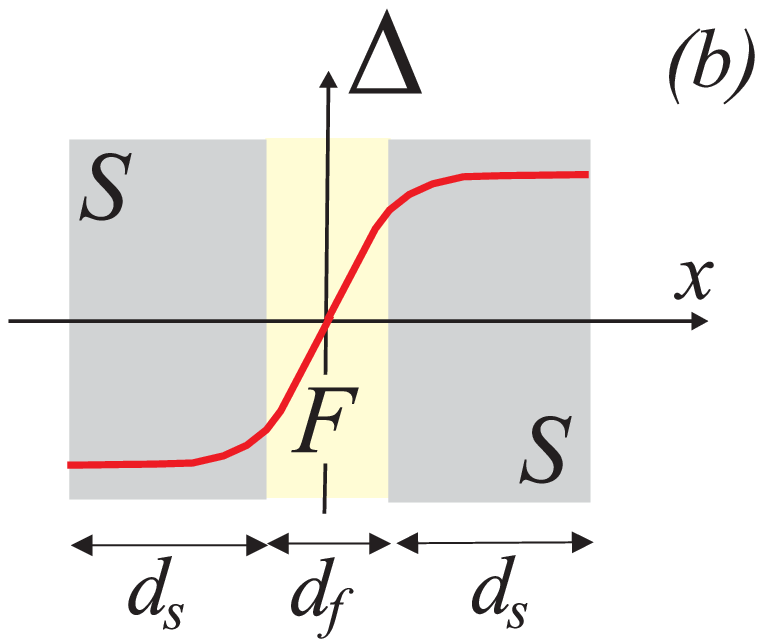}
\caption{(Color online) Schematic spatial dependence of the superconducting order parameter in S/F/S trilayers: the curve represents a sketchy behaviour of the pair wave function. (a) In the even mode ($0$--phase), due to the symmetry, the derivative of the pair wave function vanishes at the centre of the F--layer. (b) In the odd mode ($\pi$--phase), the pair wave function itself vanishes at the centre of the F--layer, and it has a $\pi$--shift in its phase in diametrically opposite points.}
%\label{Fig:1}
\label{Fig:1}
\end{figure}

In this Article we present joint experimental and theoretical results, showing the thermodynamic nature of the $0-\pi$ transition. We perform measurements of the temperature dependence of the effective London penetration depth, $\Delta\lambda_{eff}(T)$, in $\mathrm{Nb/Pd_{0.84}Ni_{0.16}/Nb}$ trilayers with relatively small $d_s \sim \xi_s$. In the sample with $d_{f} = \mathrm{2\, nm}$, close to the critical thickness, we observe a reentrant jump in the $\Delta\lambda_{eff}(T)$ curve: $\lambda_{eff}$ increases with lowering temperature, i.e. the superfluid density decreases, a strong experimental evidence of temperature mediated $0-\pi$ transition. As expected, this phenomenon  is absent in trilayers with different values of $d_f$. In agreement with the experimental data, the theoretical analysis predicts the upward  jump of $\Delta\lambda_{eff}$ as the temperature decreases providing a qualitative description of the observed phenomenon.

%---------------------------RESULTS
\section{Samples growth and experimental setup}
Nb/Pd$_{0.84}$Ni$_{0.16}$/Nb trilayers were grown on $\rm Al_2O_3$ substrates by ultrahigh vacuum  dc diode magnetron sputtering at an Ar pressure of $3\cdot10^{-6}$ Torr after obtaining a base pressure of $2\cdot10^{-8}$ Torr following the procedure described in Ref.~\onlinecite{Cirillo-SUST11}:
the samples were all prepared in the same deposition run thanks to the presence of a movable shutter in the deposition chamber which selectively covers the substrates, glued by silver paste on the holder, which were kept at room temperature during the deposition process. The typical deposition rates were 0.28 nm/s for Nb and 0.40 nm/s for Pd$_{0.84}$Ni$_{0.16}$ measured by a quartz crystal monitor previously calibrated by low-angle X-ray reflectivity measurements on deliberately deposited thin films of each material.
The Nb layers have the same nominal thickness in all the samples of the series, $d_{s}=15$ nm, while the thickness of the Pd$_{0.84}$Ni$_{0.16}$  layer changes from $d_f=2$ nm to 9 nm (the range where we expect $d_{f}\sim \xi _{f}$). The total thickness of the trilayers is then $d=2d_s+d_f$. A pure Nb film with $d=30$ nm was also grown for comparison (labelled in the following with $d_f = 0$ nm). It has a superconducting critical temperature $T_{c0}=7.5$ K and $\xi_s=\sqrt{D_{s}/2\pi T_{c0}}=6$ nm (here $D_{s}$ is the diffusion coefficient in the superconductor).\cite{Cirillo-PRB2005} Estimates of $E_{ex}\simeq$ 14 meV  and $\xi_f\simeq$ 3 nm for the $\rm Pd_{0.84}Ni_{0.16}$ alloy have been reported elsewhere\cite{Cirillo-SUST11}. The complete electrical characterization of S/F/S trilayers has been previously reported in Refs. \onlinecite{Silva-SUST11,Torokhtii-PhC12}.
High-resolution transmission electron microscopy showed excellent crystallinity of the Nb layers, roughness at the SF interfaces less than 1 nm, leading to good interface transparency, and suggested some interdiffusion of Nb into the PdNi layer.\cite{temcit} In a previous study,\cite{Pompeo-JOSC13} the local atomic structure was investigated to assess whether the F--layer could induce significant disorder in the S--layers. To this aim, extended X-ray absorption spectroscopy at the Nb K-edge was performed. The results did not show correlations between the structural disorder in the Nb layer and the superconducting properties, such as $T_c$, $H_{c2}$ and the microwave results described in the following.

Microwave measurements were performed using the dielectric resonator technique.\cite{Klein-JSup92,Torokhtii-PhC12} The quantity experimentally measured is the resonant frequency $\nu_0$ of the resonator incorporating the sample.
The sample was placed as an end wall in a cylindrical dielectric-loaded resonator. The resonant frequency of the resonator $\nu_0$ depends on the energy of the electromagnetic field stored in the volumes of the resonator and of the sample portion where the screening of the field, occurring on a length scale $\lambda_{eff}$, is not complete. Since the empty resonator gave no additional temperature dependencies in the small temperature range here explored, the resonant frequency of the resonator $\nu_0(T)$ changed only as a consequence of the change of the sample screening. Thus, the experimentally accessible quantity is the temperature variation of the effective penetration depth with respect to a given temperature $T_{ref}$, $\Delta\lambda_{eff}(T)=\lambda_{eff}(T)-\lambda_{eff}(T_{ref})$, which is obtained from $\nu_0$ through the relation:
\begin{equation}
\Delta\lambda_{eff}(T)=- \frac{G}{\pi\mu_0}\frac{\nu_0(T)-\nu_0(T_{ref})}{\nu_0^2(T_{ref})},
\end{equation}
 where $\mu_0 = 4\pi \cdot 10^{-7}\, \mathrm{H\, / \,m}$ and $G$ is a calculated geometrical factor.
 
The microwave assembly was placed in a $^4$He cryostat where temperatures down to 2.8 K were reachable. The cylindrical resonator was loaded with a Rutile (TiO$_2$) cylinder, with negligible temperature dependence of the complex permittivity below 10 K. The resonant mode chosen was the TE$_{011}$, with circular induced currents on the sample. A magnetic field up to $\mu_0H=$ 0.7 T ould be applied perpendicular to the sample plane. The setup has been extensively described previously.\cite{pompeoMSR}

Since the total thickness $d$ of the S/F/S structure is smaller than the  penetration depth, the London penetration length and the losses (see below) are averaged over the whole sample. In this full-penetration regime one has for the effective penetration depth:\cite{lambdaeff}
\begin{equation}
\lambda_{eff}(T)=\lambda(T) \coth{\frac{d}{\lambda}}\approx \frac{\lambda^2(T)}{d}.
\end{equation}
%The effective penetration depth $\lambda_{eff}(T)$ is related to $\nu_0(T)$ by $\lambda_{eff}=A-B\nu_0$, where $A$ is unknown and $B$ can be calculated. Thus, the experimentally accessible quantity is the temperature variation of the effective penetration depth with respect to a given temperature $T_{ref}$, $\Delta\lambda_{eff}(T)=\lambda_{eff}(T)-\lambda_{eff}(T_{ref})$.
The measured $\Delta \lambda_{eff}$ directly compares to the square of the superconducting parameter through $1/\lambda_{eff} \propto |\Delta|^2$. Note that due to the small thickness of the F-layer and to the fact that its conductivity is smaller than that of the Nb layers,\cite{Cirillo-SUST11} its contribution to the averaged London penetration depth is negligible.
It is also important to stress that the resonant mode used (TE$_{011}$) induces only in-plane microwave currents: no current across the SF boundaries are involved and, as a consequence, $\Delta \lambda_{eff}$ is related only to the superconducting order parameter $|\Delta|^2$, without contributions from tunnelling  between layers.
In the mixed state, the microwave currents set in motion the quantized flux lines which then contribute to the field attenuation (losses) and screening.
The simultaneous measurement of the resonator quality factor and resonant frequency allows to determine the vortex resistivity $\rho_v$, related to the forces acting on the quantized flux lines and ultimately to the vortex pinning constant $k_p$ (see below).

%the measured $\Delta \lambda_{eff}$ directly compares to the square of the superconducting parameter $1/\lambda_{eff} \sim \Delta^2$ [see also Eq.\eqref{eq:lambdaeff}]. Note that due to the small thickness of the F-layer and to the fact that its conductivity is smaller than that of the Nb layers \cite{Cirillo-SUST11}, its contribution to the averaged London penetration depth is negligible.

\section{Experimental results}
\label{results}

Raw data for the resonant frequency are reported in Fig.\,2a, for the samples with $d_f=0$, 2, and 8 nm. The flattening of $\nu_0$ close to $T_c$ is due to the loss of sensitivity when the samples become nearly electromagnetically transparent, and it is not relevant for the present purposes. The raw data show (i) that for $d_f=8$ nm the temperature dependence of the screening is smoothened with respect to pure Nb, and (ii) that the sample with $d_f=2$ nm exhibits a nonmonotonous temperature dependence of the screening. This last effect is the main experimental result of this paper, and we will concentrate on it in the following. The data of Fig.\,2a were converted to $\Delta\lambda_{eff}$, and plotted as a function of the reduced temperature, $t=T/T_c$ in Fig.\,2b. In order to have the same reference reduced temperature, $t_{ref}=T_{ref}/T_{c}=0.72$,  for all the samples, different $T_{ref}$ had to be set. It is evident that Nb ($d_f=0$) and the sample $d_f=$ 8 nm do not show any significant feature. Focussing on the sample with $d_f=2$ nm, the temperature dependence of $\Delta \lambda_{eff}$ can be divided in two regimes. Close to $T_c$ it is more similar to the one of the Nb film, while as $T$ is lowered it shows a jump upward, followed by a smoother behaviour, similar to what shown by the sample with $d_f=8$ nm. The data have not been scaled vertically: the crossover between the two different regimes is unambiguous. It is worth to stress that the upward jump of the screening with decreasing $T$ implies the {\em decrease} of the superfluid density, that is of $|\Delta|^2$, with decreasing temperature. This nonmonotonous, reentrant behaviour  is clearly the most striking result, and thus it must be thoroughly checked. The check for reproducibility was performed by repeating the measurements after disassembling and reassembling the resonator and the sample holder. While the absolute $\nu_0$ changed (as it is expected), the temperature shift yielded identical results. The results are reported in Fig.\,2c, where we show the full temperature range accessible, and they show that the two measurement sets superimpose exactly, thus excluding experimental artifacts.
%%%%%%%%%%%%%%%%%%%%%%%%%%%%%%%%%%%%%%%%%%%%
\begin{figure}[ht]
\centerline{\includegraphics[width=8 cm]{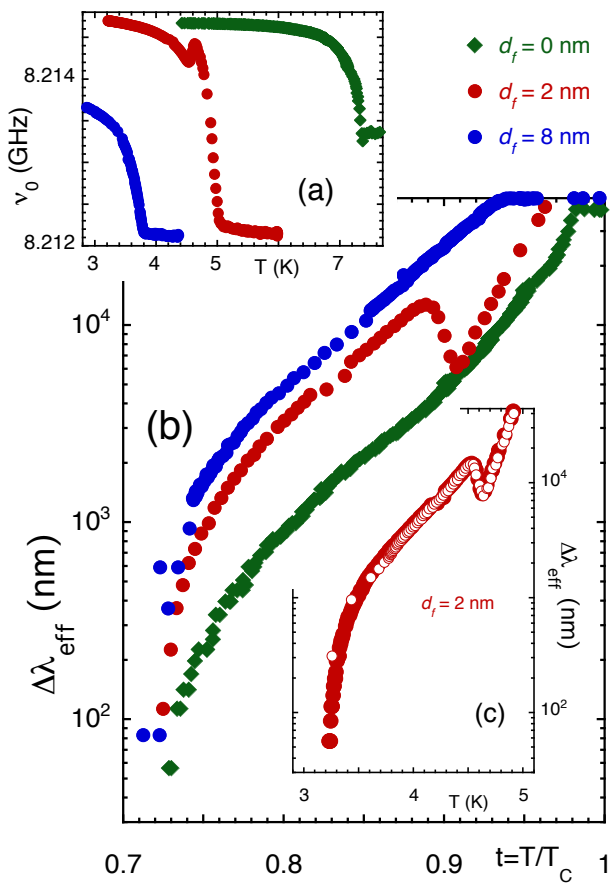}}
%\centerline{\includegraphics[width=.9\columnwidth]{Fig2.eps}}
%\vspace{-5mm}
\caption{(Color online) Temperature dependence of the screening exhibited by the trilayers:
(a) Raw data for the resonant frequency of the resonator, $\nu_0(T)$, incorporating different samples with $d_f=0$, 2, and 8 nm. The screening vanishes when the critical temperature is approached. A peak is evident in the sample with $d_f=2$ nm, showing nonmonotonous screening. The data for the sample with $d_f=8$ nm have been shifted downward by 4 MHz for clarity. The flat part at the superconducting transition is not a saturation, but the region where the resonator loses sensitivity. (b) Effective penetration depth $\Delta\lambda_{eff}(t)=\lambda_{eff}(t)-\lambda_{eff}(t_{ref})$ for the same samples, obtained with $t_{ref}=0.72$. With $d_f=2$ nm, the crossover at $t \sim 0.9$ between two different regimes, corresponding to a {\em decrease} of the superfluid fraction with decreasing temperature, is evident. (c) Reproducibility test of the peak in $\Delta\lambda_{eff}(T)$ after disassembling and reassembling the resonator (empty symbols). Full symbols, original data.}
\label{fig:raw}
%\vspace*{-5mm}
\end{figure}
%%%%%%%%%%%%%%%%%%%%%%%%%%%%%%%%%%%%%%%%%%%%
%
%%%%%%%%%%%%%%%%%%%%%%%%%%%%%%%%%%%%%%%%%%%%%
%\begin{figure}[ht]
%  % Requires \usepackage{graphicx}
%\centreline{%\includegraphics[width=1\columnwidth]{Fig23.eps}}
%%\vspace{-5mm}
%\caption{ Temperature dependence of the screening exhibited by the trilayers with
%$d_f=0$, 2, and 8 nm. Inset: raw data for the resonant
%frequency of the resonator, $\nu_0(T)$ (the data for the sample with $d_f=8$
%nm have been shifted downward by 4 MHz
%  for clarity). A peak is evident in the sample with $d_f=2$
%nm. The flat part is not a saturation, but the region where the
%resonator loses sensitivity. Main panel: $\Delta\lambda_{eff}(t)=\lambda_{eff}(t)-\lambda_{eff}(t_{ref})$
%obtained at
%$t_{ref}=0.72$. With $d_f=2$
%nm, the crossover at $t \sim 0.9$ between two different regimes,
%corresponding to a {\em decrease} of the superfluid fraction with
%decreasing temperature, is evident.}
% \label{fig:screening}
%%\vspace*{-5mm}
%\end{figure}
%%%%%%%%%%%%%%%%%%%%%%%%%%%%%%%%%%%%%%%%%%%%%
%

%\subsection{Superfluid density in the mixed state.}
%
Since the reentrant superconducting superfluid is a very relevant result, we estimated $\lambda^2\propto|\Delta|^{-2}$ in a different experiment. We applied a magnetic field $H$ perpendicular to the sample plane, thus entering the mixed state of the superconducting structure. Then, quantized flux lines are pinned by defects, which exert a recalling force (per unit length) that can be assumed to be elastic as a first approximation (small displacements $\delta r$): $F_p=-k_p \delta r$, where $k_p$ is the vortex pinning constant. Since $k_p$ ultimately depends on the energy gain of the vortex sitting on a defect, as a crude approximation $\lambda$ is connected to $k_p$ by the approximate equality between elastic and condensation energy:\cite{Golosovsky-SUST96} $\frac{1}{2}k_{p}\xi_s^2\approx c_p\frac{1}{2}\mu_0H_c^2\xi^2$ ($c_p$ is a constant of order unity). Using for the thermodynamic critical field $H_c^2\approx H_{c1}H_{c2}$ one finds:
\begin{equation}\label{eq:kappa}
    k_{p}\approx c_p H_{c2}\Phi_0/{4\pi}{\lambda^2(T)}
\end{equation}
where $\Phi_0$ is the flux quantum. The upper critical field $H_{c2}$ is obtained from dc resistivity or from the disappearance of the field-dependent microwave signal.\cite{Silva-SUST11} The pinning constant $k_p$ is easily obtained from microwave measurements of the magnetic field dependent vortex--motion complex resistivity $\rho_{v}(H)=\rho_{v1}(H)+\mathrm{i}\rho_{v2}(H)$, within a wide class of models for vortex relaxational dynamics.\cite{GR,CC,Brandt-PRL91,Pompeo-Silva-PRB08} All the details, as well as the uncertainty intervals, associated to the determination of $k_p$, have been discussed previously.\cite{Pompeo-Silva-PRB08} Thus, we can estimate the temperature dependence of the superfluid density using a conceptually different experiment, with the benefit of making the experiment {\it in situ}: we measured $\rho_{v}(H)$ at several temperatures with the same resonator described above, and we derived $k_p(H)$ within the Gittleman-Rosenblum model.\cite{GR} Numerical differences in the estimate of $k_p$ according to different models\cite{Pompeo-Silva-PRB08} are absorbed in $c_p$, and are not expected to affect the temperature dependence.
%%%%%%%%%%%%%%%%%%%%%%%%%%%%%%%%%%%%%%%%%%%%
\begin{figure}[ht]
\centerline{\includegraphics[width=8 cm]{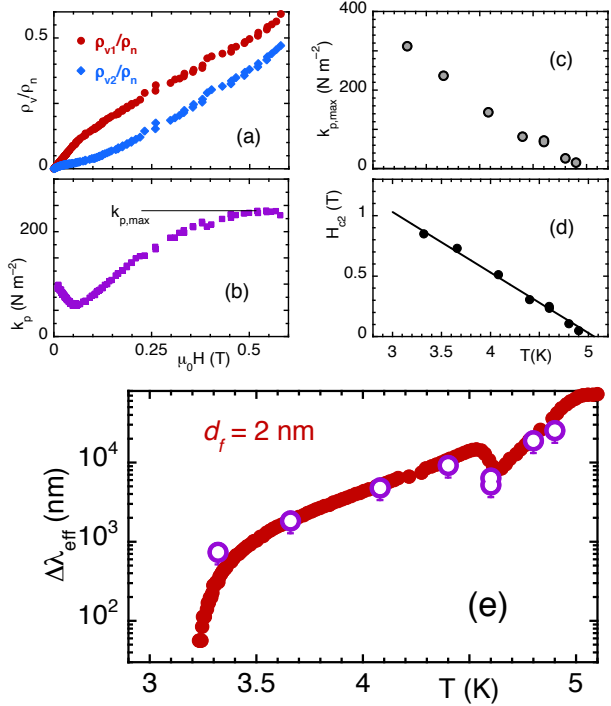}}
%\centerline{\includegraphics[width=.9\columnwidth]{Fig3.eps}}
%\vspace{-5mm}
\caption{(Color online) Confirmation of the nonmonotonous behaviour of the superfluid density from measurements in the mixed state:
(a) normalized complex vortex resistivity at $T= 3.66$ K ($\rho_n=22\, \mathrm{\mu\Omega\cdot cm}$, Ref.~\onlinecite{Torokhtii-PhC12}). (b) pinning constant $k_p$ at $T=3.66$~K, and definition of $k_{p,max}$. (c) $k_{p,max}$ as a function of $T$. (d) Upper perpendicular critical field $H_{c2}(T)$; the continuous line is a fit with $H_{c2}(0)=$ 2.53 T and $T_c=$ 5.06 K. (e) Comparison of $\Delta\lambda_{eff}(T)=\lambda_{eff}(T)-\lambda_{eff}(T_{ref})$ (here $T_{ref}=$ 3.2 K) as measured in the Meissner state (red dots) and as derived from $k_{p,max}$ (purple circles) using Eq.\eqref{eq:kappa}.}
%\vspace*{-5mm}
\end{figure}
%%%%%%%%%%%%%%%%%%%%%%%%%%%%%%%%%%%%%%%%%%%%

Fig.~3a presents sample measurements of the magnetic field dependence of the normalized vortex-state resistivity at $T=3.66$ K ($t=0.72$) for the sample with $d_f=2$ nm. The field dependence of the derived $k_p$ is reported in Fig.~3b. We have chosen to use the maximum value, $k_{p,max}$, in Eq.\eqref{eq:kappa} to derive $\Delta\lambda_{eff}$. Measurements of $k_{p,max}$ at several temperatures, reported in Fig.~3c, were thus converted into estimates for $\Delta\lambda_{eff}$ using the measured $H_{c2}(T)$, reported in Fig.~3d. The datum point for  $\Delta\lambda_{eff}$ at $T=3.66$ K exactly matches the zero-field data by taking $c_p= 1.03$, which is kept fixed. All the remaining points, plotted in Fig.~3e as large purple circles, show then a very satisfactory agreement with the zero--field data, especially taking into account the crudeness of the model.\cite{mag} This is an important check of the experimental finding, and concludes the experimental results of this paper: for the sample with $d_f=2$ nm, where the condition $d_f \sim \xi_f$ is approximately fulfilled,\cite{Cirillo-SUST11} the temperature dependence of the effective penetration depth (and thus of the superfluid density), independently probed both in the Meissner state and in the vortex state, exhibits the same nonmonotonous behaviour, with a step-like increase of the superfluid with increasing temperature at $t \sim 0.9$.

\section{Theory}

%With the support of thorough theoretical calculations, we argue in the following that the observed reentrant jump in the superfluid density is a thermodynamic manifestation of the temperature-induced $0-\pi$ transition.
With the support of thorough theoretical calculations, we argue in the following that the observed reentrant jump in the superfluid density is a thermodynamic manifestation of the temperature-induced $0-\pi$ transition. In order to explain the experimental results, we aimed at calculating the effect of the $0-\pi$ crossover on the superfluid density of an F--layer sandwiched between two  S--layers. In particular, two main issues were of interest: (i) assessing the existence of a double transition with lowering the temperature, first in the $0-$state and, at lower temperatures, in the $\pi-$state and, most important, that (ii) the superfluid density in the $\pi-$state was smaller than in the $0-$state, around the transition. The theoretical model adopted for the calculations should be able to incorporate the main experimental features.

The model is set up as follows. Since the F--layer is a weak ferromagnet we use the Usadel equations\cite{Usadel-prl70} for the averaged anomalous Green's functions $F$ and $F_{s}$ for the F-- and S--regions, respectively (see Ref.~\onlinecite{Buzdin-RMP05} for details). The presence of magnetic disorder (always present in magnetic alloys, and responsible for the main mechanism of the temperature induced $0-\pi$ transition\cite{Oboznov-Ryazanov-Buzdin-PRL06,Kushnir-EPJB,Faure-prb06}) was taken into account with the introduction of the magnetic scattering rate $\tau _{s}^{-1}$. Thin S--layers ($d_{s}\lesssim \xi _{s}$) were introduced and yielded a nearly constant order parameter in the S--layers. We assumed transparent interfaces. We expect that the main correction in case of nonideal interfaces is a reduction of the effective F thickness due to some interdiffusion processes.
%begin inserted
We first note that in practice the exchange field $h$ acting on the electron's spins in the ferromagnet and the magnetic scattering rate $\tau _{s}^{-1}$ are much larger than the superconducting critical temperature. Second, the assumption of transparent interfaces yields the boundary conditions\cite{Kuprianov-JETP88} at $x=\pm d_{f}/2$:
\begin{equation}
F_{s}=F,\, \sigma_{s}\,\partial_{x} F_{s} \vert_s =\sigma_{f}\,\partial_{x} F \vert_f,
\end{equation}
where $\sigma _{f}$ and $\sigma _{s}$ are the normal-state conductivities of the F-- and S--metals.
%end
Applying the method developed in Ref.~\onlinecite{Buzdin-PRB05}, we find the solution of the non-linear Usadel equation in the F--layer near the superconducting critical temperature obtaining the self-consistency equation for the superconducting order parameter $\Delta$,
%begin inserted
%The self-consistency equation for the superconducting order parameter $\Delta$
which in turn yields the expansion of the free energy near the critical temperature.
For the $0$-state ($\pi$-state) we should choose the even (odd) anomalous Green's functions $F$. As a result we obtain the expansion of the free energy $F^{0,\pi }(T)$ in the $0$-- or $\pi$--states (indicated by the superscript) near the critical temperature as well as the critical temperatures $T_{c}^{0,\pi }$ of the two states, and the stable state is then determined. The expansion of the free energy reads:
\begin{equation}
F^{0,\pi }(T)=E_{0}\left[ a^{0,\pi }\frac{T-T_{c}^{0,\pi }}{T_{c}^{0,\pi }}\Delta ^{2}+\frac{b^{0,\pi }}{2}\,\Delta ^{4}\right],
\end{equation}
 where $E_{0}=N(0)Ad_{f}$ is determined by the  electron density of states $N(0)$ in S--layer and by the area $A$ of the cross section of the junction, and the superscripts $0$ and $\pi$ label the quantities for the $0-$ and $\pi-$ state, respectively.
%end

The critical temperatures $T_{c}^{0,\pi }$ of the transitions into $0-$ or $\pi-$ states are given by the expressions:\cite{Buzdin-RMP05,Oboznov-Ryazanov-Buzdin-PRL06}
\begin{equation}
 \ln \left( \frac{T_{c}^{0,\pi }}{T_{c0}}\right) =    \Psi \left( \frac{1}{2} \right) -\mathrm{Re} \left\{ \Psi \left( \frac{1}{2}+\Omega _{0,\pi }\right)\right\},
 \end{equation}
where $\Psi $ is the digamma function, and $\Omega _{0,\pi }$ is the depairing parameter:
\begin{equation}
\label{eq:26}
    \Omega _{0,\pi }=\frac{\varepsilon T_{c0}}{2 T_{c}^{0,\pi }}
    \begin{cases} k\, \tanh(k s_f), & 0-\mathrm{phase}\\ k\,
        \coth(k s_f), & \pi-\mathrm{phase}
    \end{cases} \, , \nonumber
\end{equation}
with $\varepsilon=\sigma_f\xi^2_s/\sigma_sd_s\xi_f$, $s_f=d_f/2\xi_f$ and $k^{2}=2(\mathrm{i}+1/\tau _{s}h)$. The explicit
expressions for the coefficients $a^{0,\pi }$ and $b^{0,\pi }$ read:
\begin{eqnarray} \nonumber
&&a^{0,\pi } =1-\mathrm{Re}\left\{ \Omega_{0,\pi}
\Psi^{(1)}(1/2+\Omega_{0,\pi })\right\} \,, \\
\nonumber
&&b^{0,\pi } =\frac{-1}{(4\pi
T_{c}^{0,\pi})^{2}}\,\mathrm{Re}\left\{ \Psi
^{(2)}(1/2+\Omega_{0,\pi })
-\frac{\Omega_{0,\pi }}{6 k^2} \times \right.  \\
&&\quad\left.\left[i \mp \frac{i + 4/\tau_s h}{\cosh\gamma \pm
1}\left(1 \pm \frac{\gamma}{\sinh\gamma}\right)\right] \Psi
^{(3)}(1/2+\Omega _{0,\pi })\right\} \,, \nonumber
\end{eqnarray}
where $\Psi ^{(n)}(z)=d^{n}\Psi (z)/dz^{n}$, and $\gamma = 2 k s_f$.

The functional $F^{0,\pi }(T)$ provides  the complete description of the S/F/S trilayers near the critical temperature. The equilibrium energy of the system is:
\begin{equation}
F^{0,\pi }(\Delta _{0,\pi })=-E_{0}{\left[ a^{0,\pi}\,(T_{c}^{0,\pi }-T)/T_{c}^{0,\pi }\right] ^{2}}/{2b^{0,\pi }}.
\end{equation}
Thus, the first order transition between $0-$ and $\pi-$ states occurs at $F^{0}(\Delta _{0})=F^{\pi }(\Delta _{\pi })$, thus determining the transition line $T_{0}$ as:
\begin{equation}
\frac{T_{c}^{0}-T_{0}}{T_{c}^{\pi }-T_{0}}=\frac{a^{\pi }T_{c}^{0}}{%
a^{0}T_{c}^{\pi }}\sqrt{\frac{b^{0}}{b^{\pi }}}\,.
\label{eq:Tc0}
\end{equation}
The crossing of the curves $T_{c}^{0}(d_{f})$ and $T_{c}^{\pi}(d_{f})$ occurs at a value $d_{f}^{\ast }$: for $d_{f}>d_{f}^{\ast }$ it is $T_{c}^{\pi} > T_{c}^{0}$. At $d_{f}<d_{f}^{\ast}$, but at thicknesses near $d_{f}^{\ast}$, the decrease of the temperature determines first the transition from the normal to the superconducting $0-$ state and then, with the further decrease of the temperature, the transition from the $0$-- to the $\pi$--state (see inset of Fig.~4).
%%%%%%%%%%%%%%%%%%%%%%%%%%%%%%%%%%%%%%%%%%%%
\begin{figure}
\includegraphics[width=9 cm]{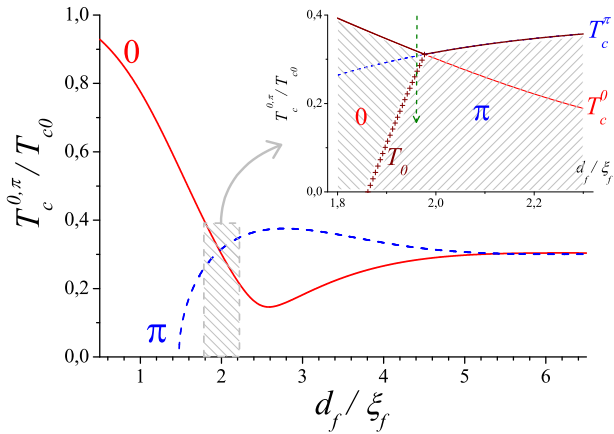}
\caption{(Color online) Critical temperatures $T_c^{0,\pi}$ vs. the thickness of the F-layer, and existence of the $0-\pi$ transition line:
the figure reports the typical dependence of $T_c^{0,\pi}$ on F-layer thickness $d_f$ for the even mode ($0-$phase, solid red line) and for odd mode ($\pi-$phase, dashed blue line). The inset zooms in the shaded region of the ($T_c,\, d_f$) phase diagram, to reveal the existence of the temperature-induced $0-\pi$ transition (dashed green vertical arrow). Symbols $+$ show the $0-\pi$ transition line $T_0(d_f)$. Here we choose: $d_s = {\rm 2} \,\xi_s$; $\sigma_f / \sigma_s = {\rm 0.12}$; $\xi_s / \xi_f = {\rm 3}$ ($\varepsilon=\mathrm{0.18}$), and $\sqrt{b^{\pi} / b^0} \simeq \mathrm{1.25}$. }
 \label{Fig:4}
\end{figure}
%%%%%%%%%%%%%%%%%%%%%%%%%%%%%%%%%%%%

Thus, we have calculated the transition temperature\cite{Buzdin-RMP05,Oboznov-Ryazanov-Buzdin-PRL06} in the $0-$ and $\pi-$states, as a function of the ratio $d_f/\xi_f$, $T_c^{0,\pi}(d_f/\xi_f)$, and we have found and calculated the transition line between the $0-$ and the $\pi-$states, $T_0(d_f/\xi_f)$. The typical $T_{c}^{0,\pi }(d_{f})$ dependence is presented in Fig.~4. In this diagram, one must stress that near the critical thickness, where $T_{c}^{0 }=T_{c}^{\pi }$, the $0-\pi$ transition line $T_0$ emerges when $F^{0}(\Delta _{0})=F^{\pi }(\Delta _{\pi })$. Only when $d_f$ is below, but close to, the critical thickness, the decrease of the temperature determines first the transition from the normal to the superconducting $0$-state and then, with the further decrease of the temperature, the transition from the $0$-to the $\pi$-state, as shown in the inset of Fig.~4.

%At the transition, see Fig.~5a, the superconducting order parameter jumps from $\Delta _{0}$ to $\Delta _{\pi }$ according to $\Delta _{\pi }^{2}(T_{0})=\frac{a^{\pi }}{b^{\pi }}\left({T_{c}^{\pi}-T_{0}}\right)/T_{c}^{0} =\Delta _{0}^{2}(T_{0})\sqrt{{b^{0}}/({b^{\pi }})}$.

To complete the picture of the temperature induced $0-\pi$ transition, and to find the quantity directly  observed in experiments, we calculate the superconducting order parameter as a function of the (lowering) temperature. In fact, at the transition, see Fig.~5a, the superconducting order parameter jumps from $\Delta _{0}$ to $\Delta _{\pi }$ according to:
\begin{equation}
\Delta _{\pi }^{2}(T_{0})=\frac{a^{\pi }}{b^{\pi }}\left(\frac{T_{c}^{\pi}-T_{0}}{T_{c}^{0}}\right) =\Delta _{0}^{2}(T_{0})\sqrt{\frac{b^{0}}{b^{\pi }}}.
\end{equation}
 It is an essential result that, since $b^0/b^{\pi}< 1$  in the whole range of reasonable values for the various parameters (see Fig.~5c), at the transition the order parameter in the $\pi-$state has a {\em smaller} value than in the $0-$state, whence the reentrant behaviour of the superfluid density and the upward jump of the London penetration depth $1 / \lambda^2 \sim \Delta ^{2}$, see Fig.~5b. So our theoretical description completely recover the experimental findings of Section \ref{results}.
%%%%%%%%%%%%%%%%%%%%%%%%%%%%%%%%%%%%%%%%%%%%
\begin{figure}[t]
\includegraphics[width=8.5 cm]{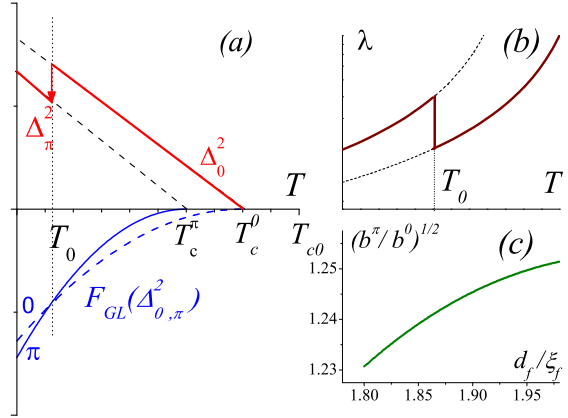}
\caption{ (Color online) Theoretical temperature dependence of the Ginzburg--Landau parameters of the S/F/S structure.
(a) Schematic temperature dependence of the gap $\Delta_{0,\pi}^2$ (upper scale, red lines) and the Ginzburg--Landau energy $F_{GL}(\Delta_{0,\pi}^2)$ (lower scale, blue lines): $T_c^{\pi} < T_c^{0}$ and $(a^{\pi})^2/b^{\pi} > (a^{0})^2/b^{0}$; (b) schematic temperature dependence of the penetration depth $\lambda \sim 1 / \Delta_{0,\pi}$; (c) dependence of the superconducting gap jump $\Delta_0^2(T_0)/\Delta_{\pi}^2(T_0)=(b_\pi/b_0)^{1/2}$ on the F-layer thickness $d_f \lesssim d_f^*$. The parameters are the same as in Fig.~4.}
\label{Fig:5}
\end{figure}
%%%%%%%%%%%%%%%%%%%%%%%%%%%%%%%%%%%%

\section{Conclusion}
\label{conc}

In summary, we have shown for the first time the thermodynamic nature of the $0-\pi$ transition, by monitoring the temperature dependence of the order parameter close to the critical thickness condition. The experimental demonstration relied on the observation, by two different experiments, of a reentrant jump in the Cooper pair density in the Meissner and in the mixed states. The observation was performed in an S/F/S heterostructure with the F--layer close to the critical thickness, and with thin S--layers. An accurate theoretical treatment allowed us to find and calculate the $0-\pi$ transition line, as well as the existence of the reentrant jump in the superfluid density at the temperature-induced $0-\pi$ transition. We have concluded that the observed jump  was related to the first order transition from 0- to $\pi$-state on cooling.

\section*{Acknowledgments}
The authors thank A.S. Mel'nikov and C. Meneghini for stimulating discussions, and S. Sarti and R. Loria for their help in the microwave measurements. This work was supported, in part, by European NanoSC COST Action
MP1201, by French ANR grant ``MASH'', by the RFBR (grant n.13-02-97126), by the program ``Quantum Mesoscopic and Disordered Structures'' of the RAS, and by Ministry of Education and Science of RF and Lobachevsky State University (agreement 02.B.49.21.0003).

\end{document}